\documentclass[a4paper,11pt]{article}
\usepackage{graphicx}
\usepackage{cite}
\usepackage{hyperref}

\date{\today}

\usepackage{color}

\usepackage{xspace}
\newcommand{\kks}{KBHsSH\xspace}
\newcommand{\kk}{KBHSH\xspace}

\usepackage{amsmath}
\usepackage{amssymb}
\usepackage[hang,nooneline,scriptsize]{subfigure}
\begin{document}

\title{{\bf  \LARGE Astrophysical imaging  \\ of Kerr black holes with scalar hair}}

\author{
{\large F. H. Vincent}$^{1}$,
{\large E. Gourgoulhon}$^{2}$,
{\large C. Herdeiro}$^{3}$
and
{\large E. Radu}$^{3}$
\\
\\
$^{1}${\small LESIA, CNRS UMR 8109, Observatoire de Paris} \\{\small Universit\'e Pierre et Marie Curie, Universit\'e Paris Diderot}\\{\small 5 place Jules Janssen, 92190 Meudon, France}
 \\ \texttt{\small frederic.vincent@obspm.fr}
\\
\\
$^{2}${\small LUTH, CNRS UMR 8102, Observatoire de Paris } \\ {\small Universit\'e Paris Diderot} \\{\small 5 place Jules Janssen, 92190 Meudon,
France}
 \\ \texttt{\small eric.gourgoulhon@obspm.fr}
\\
\\
$^{3}${\small Departamento de F\'\i sica da Universidade de Aveiro and  } \\ {\small  Centre for Research and Development  in Mathematics and Applications (CIDMA),
 } \\ {\small    Campus de Santiago, 3810-183 Aveiro, Portugal}
 \\ \texttt{\small  herdeiro@ua.pt; eugen.radu@ua.pt}
}

%

\date{June 2016}
\setlength{\footnotesep}{0.5\footnotesep}
\newcommand{\tr}{\mbox{tr}}
\newcommand{\la}{\lambda}
\newcommand{\ka}{\kappa}
\newcommand{\f}{\phi}
\newcommand{\vf}{\varphi}
\newcommand{\F}{\Phi}
\newcommand{\al}{\alpha}
\newcommand{\ga}{\gamma}
\newcommand{\de}{\delta}
\newcommand{\si}{\sigma}
\newcommand{\bomega}{\mbox{\boldmath $\omega$}}
\newcommand{\bsi}{\mbox{\boldmath $\sigma$}}
\newcommand{\bchi}{\mbox{\boldmath $\chi$}}
\newcommand{\bal}{\mbox{\boldmath $\alpha$}}
\newcommand{\bpsi}{\mbox{\boldmath $\psi$}}
\newcommand{\brho}{\mbox{\boldmath $\varrho$}}
\newcommand{\beps}{\mbox{\boldmath $\varepsilon$}}
\newcommand{\bxi}{\mbox{\boldmath $\xi$}}
\newcommand{\bbeta}{\mbox{\boldmath $\beta$}}
\newcommand{\bea}{\begin{eqnarray}}
\newcommand{\eea}{\end{eqnarray}}
\newcommand{\be}{\begin{equation}}
\newcommand{\ee}{\end{equation}}

\newcommand{\nn}{\nonumber}
\newcommand{\mm}{\mathcal{M}}
\newcommand{\dd}{\mathrm{d}}
\newcommand{\pp}{\varphi}

\newcommand{\ii}{\mbox{i}}
\newcommand{\e}{\mbox{e}}
\newcommand{\pa}{\partial}
\newcommand{\Om}{\Omega}
\newcommand{\vep}{\varepsilon}
\newcommand{\bfph}{{\bf \phi}}
\newcommand{\lm}{\lambda}
\def\theequation{\arabic{equation}}
\renewcommand{\thefootnote}{\fnsymbol{footnote}}
\newcommand{\re}[1]{(\ref{#1})}
\newcommand{\R}{{\rm I \hspace{-0.52ex} R}}
\newcommand{\N}{{\sf N\hspace*{-1.0ex}\rule{0.15ex}%
{1.3ex}\hspace*{1.0ex}}}
\newcommand{\Q}{{\sf Q\hspace*{-1.1ex}\rule{0.15ex}%
{1.5ex}\hspace*{1.1ex}}}
\newcommand{\C}{{\sf C\hspace*{-0.9ex}\rule{0.15ex}%
{1.3ex}\hspace*{0.9ex}}}
\newcommand{\eins}{1\hspace{-0.56ex}{\rm I}}
\renewcommand{\thefootnote}{\arabic{footnote}}
 \maketitle
\begin{abstract}
We address the astrophysical imaging of a family of deformed Kerr black holes (BHs). These are  stationary, asymptotically flat black hole (BH) spacetimes, that are solutions of General Relativity minimally coupled to a massive, complex scalar field: Kerr BHs with scalar hair (\kks). Such BHs bifurcate from the vacuum Kerr solution and can be regarded as a horizon within a rotating boson star. In a recent letter~\cite{Cunha:2015yba}, it was shown that \kks can exhibit very distinct shadows from the ones of their vacuum counterparts. The setup therein, however, considered the light source to be a celestial sphere sufficiently far away from the BH. Here, we analyse \kks surrounded by an emitting torus of matter, simulating a more realistic astrophysical environment, and study the corresponding lensing of light as seen by a very far away observer, to appropriately model ground-based observations of Sgr A$^*$. We find that the differences in imaging between \kks and comparable vacuum Kerr BHs remain, albeit less dramatic than those observed for the corresponding shadows in the previous setup.
In particular, we highlight two observables that might allow differentiating \kks and Kerr BHs.
The first is the angular size of the photon ring (in a Kerr spacetime) or lensing ring (in a \kk spacetime), the latter being significantly smaller
for sufficiently non-Kerr-like spacetimes.
The second is the existence of an edge in the intensity distribution (the photon ring in Kerr spacetime). This edge can disappear for very non-Kerr-like \kks.
It is plausible, therefore, that sufficiently precise Very Long Baseline Interferometric observations of BH candidates can constrain this model.
\end{abstract}
\medskip
\medskip



\section{Introduction}
Over the last half a century, electromagnetic observations have gathered a solid body of evidence for the existence of astrophysical black holes (BHs) -- see $e.g.$~\cite{Narayan:2013gca}. Recently, the detection of the gravitational wave event GW150914~\cite{Abbott:2016blz}, has strengthened the case for astrophysical BHs, now using a completely independent channel. Observations with ever increasing precision, both in the gravitational wave and electromagnetic channels, are expected over the next decade, leading to a more thorough understanding of the true nature of astrophysical BH candidates and, in particular, discriminating the paradigmatic BHs of General Relativity (GR) from more exotic alternatives~\cite{Berti:2015itd}. It is thus quite timely to obtain the appropriate phenomenology for any theoretically sound alternative model.

Steady-state BHs in vacuum GR have been recognised  to be surprisingly simple since the 1970s: they have only two degrees of freedom~\cite{Carter:1971zc,Hawking:1972,Chrusciel:2012jk} and are described by the elegant Kerr metric~\cite{Kerr:1963ud}.\footnote{Here we are taking the viewpoint that astrophysical BHs are essentially electrically neutral.} This geometry provides the standard model to obtain astrophysical phenomenology for BH candidates. Yet, assuming \textit{a priori} the Kerr metric, introduces an undesirable theoretical bias. Indeed, the Kerr solution is unique in vacuum, but different BH solutions may exist when considering generic types of matter (or departing from GR). 
{Three main avenues
are being investigated to avoid this bias. The first} is to study geometries where one introduces, \textit{ad hoc}, deformations of the Kerr geometry~(see $e.g.$~\cite{Johannsen:2013rqa,Cardoso:2014rha,Ghasemi-Nodehi:2016wao}), and investigate how much observations can constrain such deformations. 
{A second avenue corresponds to attempts of studying generic black-hole spacetimes under some sufficiently broad parameterization scheme, with Schwarzschild and Kerr being particular realizations of the general framework, rather than fiducial solutions~\cite{Rezzolla:2014mua, Konoplya:2016jvv}}.
{These two first approaches} may be informative, but {have} the unsatisfactory feature that the deformed geometries, in general, do not have a clear origin as solutions of any sensible model.  A {third} and more satisfactory first-principles approach, on the other hand, $i.e.$ starting from a concrete GR plus a (physically reasonable and astrophysically plausible) matter model, is hampered by the obvious difficulty in finding qualitatively new BH solutions, let alone solutions which are deformations of the vacuum Kerr BH.

Unexpectedly,  in the last two years, new families of deformed Kerr BHs could be obtained within simple matter models. The first (and simplest) such family that was discovered, corresponds to Kerr BHs with scalar hair (\kks)~\cite{Herdeiro:2014goa} (see also~\cite{Herdeiro:2014ima,Herdeiro:2015gia}). These are solutions of Einstein's gravity minimally coupled to a massive, complex, scalar field. This matter content obeys all energy conditions, and the solutions are regular on and outside an event horizon, thus providing a theoretically consistent model. Besides the vacuum limit, wherein they reduce to a subset of vacuum Kerr BHs, \kks have also a solitonic limit, in which they reduce to well-known gravitating solitons known as \textit{boson stars}~\cite{Schunck:2003kk,Liebling:2012fv}. \kks have been found numerically, but a formal proof of their existence was given in~\cite{Chodosh:2015oma}. Generalizations with self-interactions were constructed in~\cite{Kleihaus:2015iea,Herdeiro:2015tia} and with massive Proca fields in~\cite{Herdeiro:2016tmi}.

\kks inherit a well-known property from boson stars: the maximal ADM (Arnowitt, Deser, Misner) mass possible is of the order of the Compton wavelength of the scalar field. This demands, for the existence of either boson stars or \kks  with ADM masses of the order of the solar mass, or higher, that the scalar field must be ultra-light, with a mass of order of $10^{-10}$ eV, or smaller  -- see the discussion in~\cite{Herdeiro:2015tia} for models including self-interactions. Such ultralight particles are not present in the standard model of particle physics, but have been predicted in beyond standard model scenarios~\cite{Arvanitaki:2009fg}. Thus, evidence for boson stars or \kks could be faced as evidence for an ultralight particle beyond the standard model. Claiming such evidence, however, will only be possible if the phenomenology of these exotic compact objects is understood in detail and smoking guns that distinguish them from standard compact objects are identified.

Within the electromagnetic channel, a promising class of observations pertains to the lensing of light in the neighbourhood of some supermassive BH candidates, together with their \textit{shadows}~\cite{Falcke:1999pj}. 
%
{The shadow of a BH is the region in the observer's sky comprising the directions of photons that asymptotically approach the event horizon in a backward ray-tracing computation.}
This type of observable is being targeted by the Event Horizon Telescope (EHT)~\cite{2009astro2010S..68D}, a millimeter-wavelength Very Long Baseline Interferometry (VLBI) network, which has the potential to test the Kerr paradigm and constrain alternative models~\cite{2014ApJ...784....7B}.
{In particular, the EHT will be able to image the shadow of the supermassive compact object at the center of the Galaxy, Sgr~A*. The resolution of the instrument will
reach $\approx 20\,\mu$as, which
is smaller than the typical size of Sgr~A*'s shadow, $\approx 50\,\mu$as}.

In a recent letter~\cite{Cunha:2015yba}, it was shown that the shadows of \kks can be sharply distinct from the ones of the vacuum Kerr BHs, in some regions of the parameter space. This study, however, used a setup (first introduced in~\cite{Bohn:2014xxa}) which had a primarily goal of producing visually striking images of the lensing, and the corresponding differences with respect to Kerr, rather than simulating a realistic astrophysical environment. Thus, a natural question is whether introducing a more realistic astrophysical environment masks the peculiar shadows obtained in~\cite{Cunha:2015yba}. The main purpose of the current paper is to address this question.

We shall consider the same configurations of \kks studied in detail in~\cite{Cunha:2015yba} and study their lensing, but with three main differences with respect to the former study. Firstly, the ray tracing will be done with a well tested code (\textsc{Gyoto}~\cite{Vincent:2011wz}), completely independent  from that used in~\cite{Cunha:2015yba}. Secondly, the BHs are surrounded by an accretion torus, using the same model studied recently around Kerr BHs~\cite{vincent15kerr} and boson stars~\cite{Vincent:2015xta}. This is the light source, rather than a faraway celestial sphere, as in the setup considered in~\cite{Cunha:2015yba}. Thirdly, we consider the observer to be at a realistic distance to model a
ground-based observation of Sgr A$^*$, whereas the observations collected in~\cite{Cunha:2015yba} were from the viewpoint of a much closer observer.  Our main conclusion is that, even with this more realistic astrophysical setup, observable differences remain between the hairy BHs and their vacuum counterparts, even though the astrophysical environment partly masks the strikingly different shapes observed in~\cite{Cunha:2015yba}. More concretely, the total flux difference in the imaging of \kks and comparable Kerr BHs is small, for the sample of cases analysed; but the photon ring size difference can have an observable signature. In more extreme cases the shadow, in the astrophysical setup, is essentially erased, in sharp contrast to that of the comparable Kerr BH.

This paper is organized as follows. In Section~\ref{sec2} we briefly review \kks and, in particular, the sample of backgrounds to be addressed in this paper. In Section~\ref{sec3} we benchmark the use of \textsc{Gyoto} by comparing its results with the ones previously obtained for the shadows of \kks. In Section~\ref{sec4} we introduce the astrophysical environment around \kks and produce the corresponding imaging. We conclude in Section~\ref{sec5} with a summary of our results and an outlook.

\section{The black hole backgrounds}
\label{sec2}
The simplest \kks are solutions of the Einstein-Klein-Gordon model, where the scalar field is free (no self-interactions), complex and massive (see~\cite{Kleihaus:2015iea,Herdeiro:2015tia} for generalizations including self-interactions). This model is described by the action:
 \begin{equation}
\label{actionscalar}
S=\int  d^4x \sqrt{-g}\left[ \frac{R}{16\pi G}
   -\frac{g^{\alpha\beta}}{2} \left( \Phi_{, \, \alpha}^* \Phi_{, \, \beta} + \Phi _
{, \, \beta}^* \Phi _{, \, \alpha} \right) - m^2 \Phi^*\Phi
 \right]  \ .
\end{equation}
Throughout we use units with $c=1=\hbar$. This model has two constants, Newton's constant $G$ and the scalar field mass $m$. We take the two associated natural scales to be Planck's mass $m_P=G^{-1/2}$, and the mass scale for boson stars/\kks, which reads
\begin{equation}
\mathcal{M} = \frac{m_P^2}{m} \ .
\end{equation}
Indeed, the maximal ADM mass for boson stars/\kks is $\alpha \mathcal{M}$, where $\alpha$ is a constant of order unity that depends on the particular type of boson stars -- see~\cite{Herdeiro:2015tia} and \cite{Grandclement:2014} for a sample of concrete $\alpha$ values.

The hairy Kerr BH solutions are found with the following metric ansatz
for a stationary axisymmetric and circular spacetime:
\begin{eqnarray}
 ds^2=-e^{2F_0}N dt^2+e^{2F_1}\left(\frac{dr^2}{N}+r^2 d\theta^2\right) + e^{2F_2}r^2 \sin^2\theta \left(d\varphi-W dt\right)^2 \ ,
 \label{kerrnc}
\end{eqnarray}
where
\begin{equation}
N\equiv 1 -\frac{r_H}{r} \ ,
\label{n}
\end{equation}
$r_H$ being a constant (representing the radial coordinate of the event horizon),
and  $F_1$, $F_2$ and $W$ are functions of the spheroidal coordinates $(r,\theta)$. The vacuum Kerr metric can be written in this coordinate system. The explicit form of the coefficients can be found in~\cite{Herdeiro:2015gia,Herdeiro:2016tmi}. Note that the parameters $b$ (in~\cite{Herdeiro:2016tmi}) and $c_t$ (in~\cite{Herdeiro:2015gia}) relate as $b=-c_t$. In the following we shall dub these as \textit{spheroidal prolate (SP) coordinates}, $cf.$ Appendix A of~\cite{Herdeiro:2016tmi}.

\kks form a countable number of families, labeled by the azimuthal harmonic index and the number of nodes in the scalar field. Here we shall focus on a particular member of this family, with the lowest azimuthal harmonic index (equal to one) and the lowest number of nodes (no nodes -- see~\cite{Herdeiro:2015gia} for a discussion of the general case). The latter defines fundamental states; excited solutions are likely unstable, towards decay into fundamental solutions, as in the case of boson stars~\cite{Balakrishna:1997ej}. The corresponding scalar field ansatz is
\begin{equation}
\Phi(t,r,\theta,\phi)=e^{-iwt}e^{i\varphi}\phi(r,\theta) \ ,
\label{sfansatz}
\end{equation}
where $w$ is a constant.
In the following we shall address three particular solutions of the model~\eqref{actionscalar}, with the ansatz~\eqref{kerrnc}--\eqref{sfansatz}, which are dubbed Configurations I-III in~\cite{Cunha:2015yba}. The numerical data for these solutions is publicly available~\cite{datakbhsh}.  Table~\ref{tab:configurations} provides a brief description of the physical parameters of these configurations (from~\cite{Cunha:2015yba}).

\begin{table}[htbp!]
\centering \caption{\kks configurations considered in the present study.
$M{}$ is the ADM mass, $M_{\rm H}$ is the horizon's Komar mass,
$J$ is the total Komar angular momentum and
$J_{\rm H}$ is the horizon's Komar angular momentum.}
\vspace{0.2cm}
\begin{tabular}{l*{7}{c}r}
\hline
  & $M{}$ & $M_{\rm H}$ & $J{}$ & $J_{\rm H}$ & $ \frac{M_{\rm H}}{M{}}$ & $\frac{J_{\rm H}}{J{}}$&$\frac{J{}}{M{}^2}$ & $\frac{J_{\rm H}}{M_{\rm H}^2}$\\
\hline
{\bf Configuration I}  & 0.415$\mathcal{M}$ & 0.393$\mathcal{M}$ & 0.172$\mathcal{M}^2$ & 0.150$\mathcal{M}^2$ & 95\% & 87\% & 0.999 & 0.971\\
 \hline
{\bf Configuration II} & 0.933$\mathcal{M}$ & 0.234$\mathcal{M}$ & 0.740$\mathcal{M}^2$ & 0.115$\mathcal{M}^2$ & 25\% & 15\% & 0.850 & 2.10\\
 \hline
{\bf Configuration III} & 0.975$\mathcal{M}$ & 0.018$\mathcal{M}$ & 0.85$\mathcal{M}^2$ & 0.002$\mathcal{M}^2$ & 1.8\% & 2.4\% & 0.894 & 6.20\\
 \hline
\end{tabular}\\
\label{tab:configurations}
\end{table}

{These three \kks configurations are compared to Kerr BH solutions with the same values of ADM
mass and total angular momentum. These comparable Kerr solutions are denoted Kerr$_{=\mathrm{ADM}}$ in~\cite{Cunha:2015yba}.
They are comparable observation-wise, because observations typically give access to the
parameters $M{}$ and $J{}$.}

Configuration I is a ``rather Kerr-like" \kk. Only $5\%$ of the mass and $13\%$ of angular momentum are stored in the scalar field. The horizon is also Kerr-like, in the sense that the Kerr bound is obeyed in terms of horizon quantities -- a property which is not mandatory for other \kks~\cite{Herdeiro:2015moa}. Its  Kerr-like shadow's average radius is only a few percent smaller than that of {its comparable Kerr counterpart.}
The latter is an almost extremal BH, with $j{}\equiv {J{}}/{M{}^2}=0.999$.

Configuration II is a ``not-so-Kerr-like" \kk. In this case, the majority of the mass ($75\%$) and angular momentum ($85\%$) are stored in the scalar field. The horizon is non-Kerr-like, in the sense that it violates the Kerr bound in terms of horizon quantities. Its shadow is not only 25$\%$ smaller than that of the comparable Kerr BH~\cite{Cunha:2015yba}, but it has also a peculiar shape -- more square -- than that observed for any vacuum Kerr BH. The comparable Kerr BH is less extremal than the corresponding one for configuration I, with $j{}=0.85$.

Finally, configuration III, is a ``very non-Kerr-like" \kk. Almost all mass ($98.2\%$) and angular momentum ($97.6\%$) are stored in the scalar field. The horizon is very non-Kerr-like, violating the Kerr bound in terms of horizon quantities by a factor of 6. The horizon shape is quite exotic -- it is delimeted by a non-convex curve and there are multiple disconnected shadows. The comparable Kerr BH has $j{}=0.894$.

In the following two sections we will readdress these configurations, firstly recomputing their shadows using \textsc{Gyoto}, and comparing the results with those previously obtained, and then performing their imaging in the astrophysical setup.

%


\section{Ray-tracing setup and shadow computations}
\label{sec3}

We use in this article the ray-tracing code \textsc{Gyoto}, which is open-source
software~\cite{gyoto,Vincent:2011wz}. We employ this code to integrate numerically null geodesics in different \kks numerical spacetimes and integrate the radiative transfer
equation inside an accretion structure surrounding the BH. The geodesic integration is performed backward in time from
a distant observer. In our setup, the observer is located at a radial coordinate corresponding to the distance between the
Earth and the Galactic Center, i.e. $8.33~$kpc~\cite{gillessen09}.
The \kks and Kerr solutions that we use are expressed in SP coordinates $(t,r,\theta,\varphi)$, as defined in~\cite{Herdeiro:2015gia}.
In order to relate the SP radial coordinate
$r$, expressed in units of $\mathcal{M}$, to a physical distance of $8.33~$kpc, we need to fix the mass of Sgr~A*.
Throughout this article we use $M = 4.31\times10^6\,M_\odot$~\cite{gillessen09}, where
$M$ is the ADM mass of the BH.
The integration is performed using a
Runge-Kutta-Fehlberg
adaptive-step integrator at order 7/8 as implemented in the \texttt{boost} C++ library.
A recent study~\cite{grould16} has demonstrated the ability of \textsc{Gyoto} to ray trace accurately even
over such very large distances.
The observer is located at some fixed inclination $\theta=85^\circ$, where $\theta$ is the
angle between the axis of rotation and the observer. This particular value of inclination is inherited from
a previous study~\cite{vincent15kerr} where it was shown to be able to reproduce well the observed
features of Sgr~A* modeled as a Kerr BH.
The observer is modeled by a screen, with every pixel corresponding to some direction
of photon incidence. The total computed field of view is typically of $300\,\mu$as,
unless otherwise stated.
The various pixels of the screen are assigned with the value
of the specific intensity transported by the photon corresponding to the pixel's direction of incidence.
The output of the ray-tracing calculation is thus a map of specific intensity over some small
field of view, which, in the following, will be called \textit{an image}.

We use the numerical metrics corresponding to the \kks and comparable Kerr solutions described in the previous
section, which are publicly available~\cite{datakbhsh}. We extended the \textsc{Lorene} library~\cite{lorene} to make it able
to read these metrics and translate them to multi-domain spectral grids. \textsc{Gyoto} is then able
to perform ray tracing using such numerical spacetimes. The
\textsc{Lorene} class \texttt{ScalarBH} used for generating \textsc{Gyoto}-compatible metrics from the \kks raw data is publicly available
in the latest version of \textsc{Lorene}. {We remark that in preparing the framework for this study, we have compared rotating boson star solutions
numerically generated by the \textsc{Kadath} code \cite{kadath} in
\cite{Grandclement:2014} with those generated by
the code used in~\cite{Herdeiro:2014goa},  \textsc{Fidisol/Cadsol}~\cite{schoen}, finding a very good agreement.}



\bigskip

In this section, we want to compute the shadows of the three \kks, configurations I, II and III, and their comparable Kerr BHs. Our aim is first to determine what the shadows
look like in an ``astrophysically neutral" setup ($i.e.$ no emission of electromagnetic radiation), which will be useful for discussing
the astrophysically realistic images later on. We also want to compare our computations with the previous
calculations developed in~\cite{Cunha:2015yba}, to ensure consistency between the two
completely independent ray-tracing codes. As a consequence, we do not consider any
source of radiation and simply trace null geodesics until they approach the event horizon.
The ray-traced images show only two intensity values: either $1$
when the backtraced photon came arbirary close to the event horizon, or $0$
otherwise.

\begin{figure}[htbp]
\centering
\includegraphics[width=6cm,height=6cm]{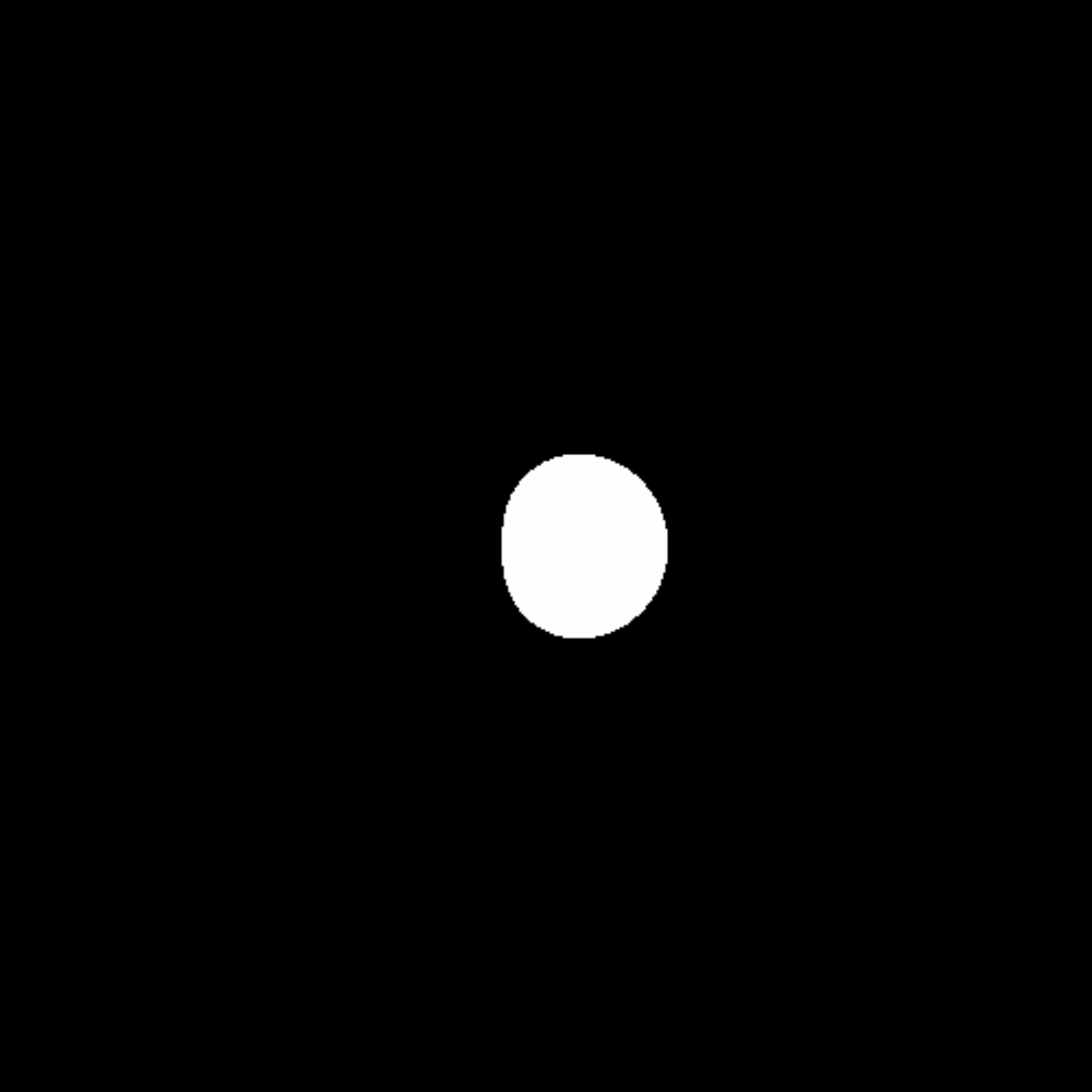}
\includegraphics[width=6cm,height=6cm]{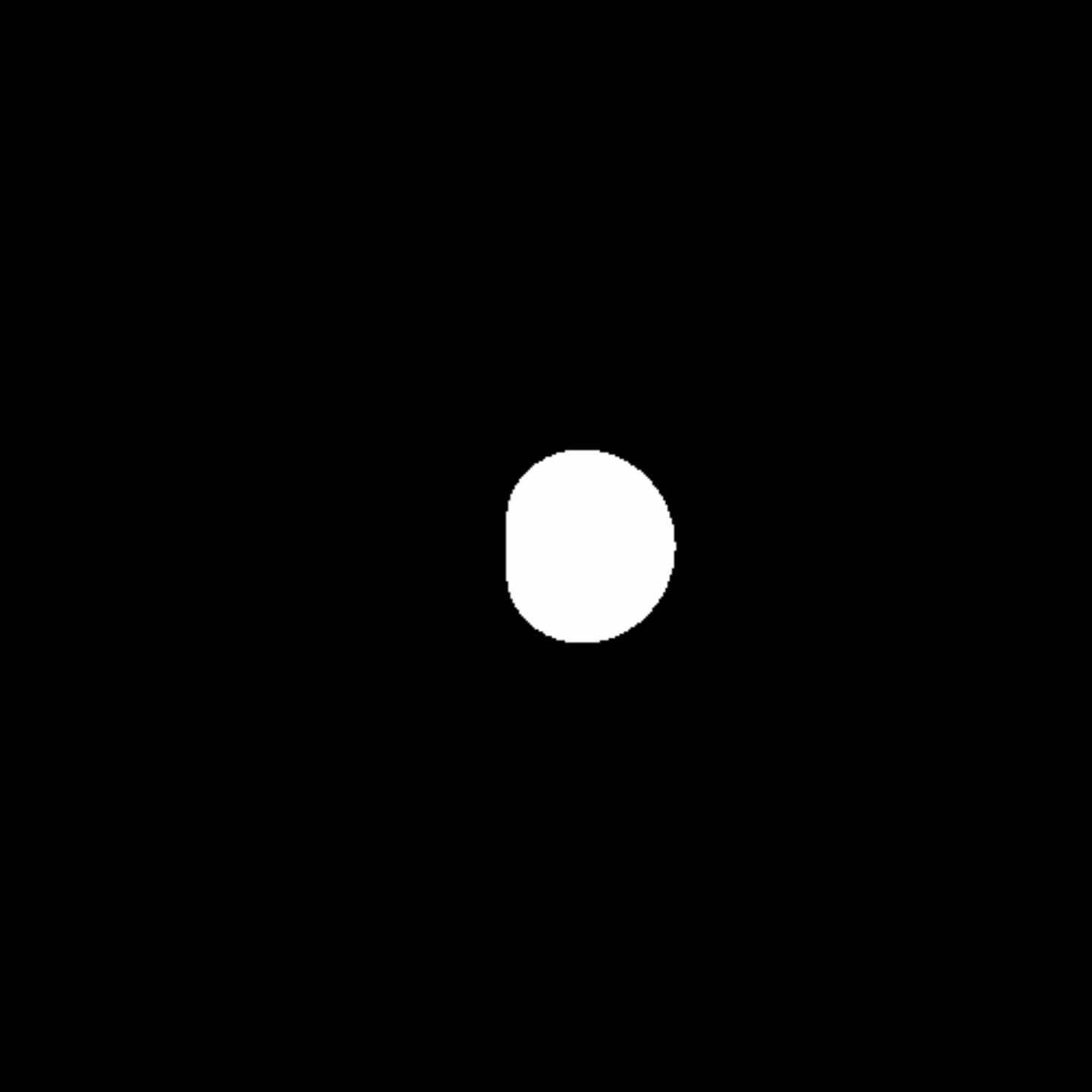}
\caption{\textit{\textbf{Left:} shadow of the \kk of configuration I (field of view $300\,\mu$as).
\textbf{Right:} the same for the comparable Kerr BH ($j_{\mathrm{ADM}}=0.999$).}}
\label{fig:shadow1}
\end{figure}

Fig.~\ref{fig:shadow1} shows the shadows of the \kk and comparable Kerr BH of our configuration I.
They look very similar to Fig.~5, top-left and top-middle panels of~\cite{Cunha:2015yba}.
Note, however, that therein an inclination of $\theta=90^\circ$ was used, whereas here we take
$\theta=85^\circ$. The inclination impact is very mild, for this configuration,
and the change in $\theta$ is barely noticible in the shadow. {Observe that in all solutions presented in this paper the BHs are rotating such that the left hand side of the image is moving towards the reader.}

\begin{figure}[htbp]
\centering
\includegraphics[width=6cm,height=6cm]{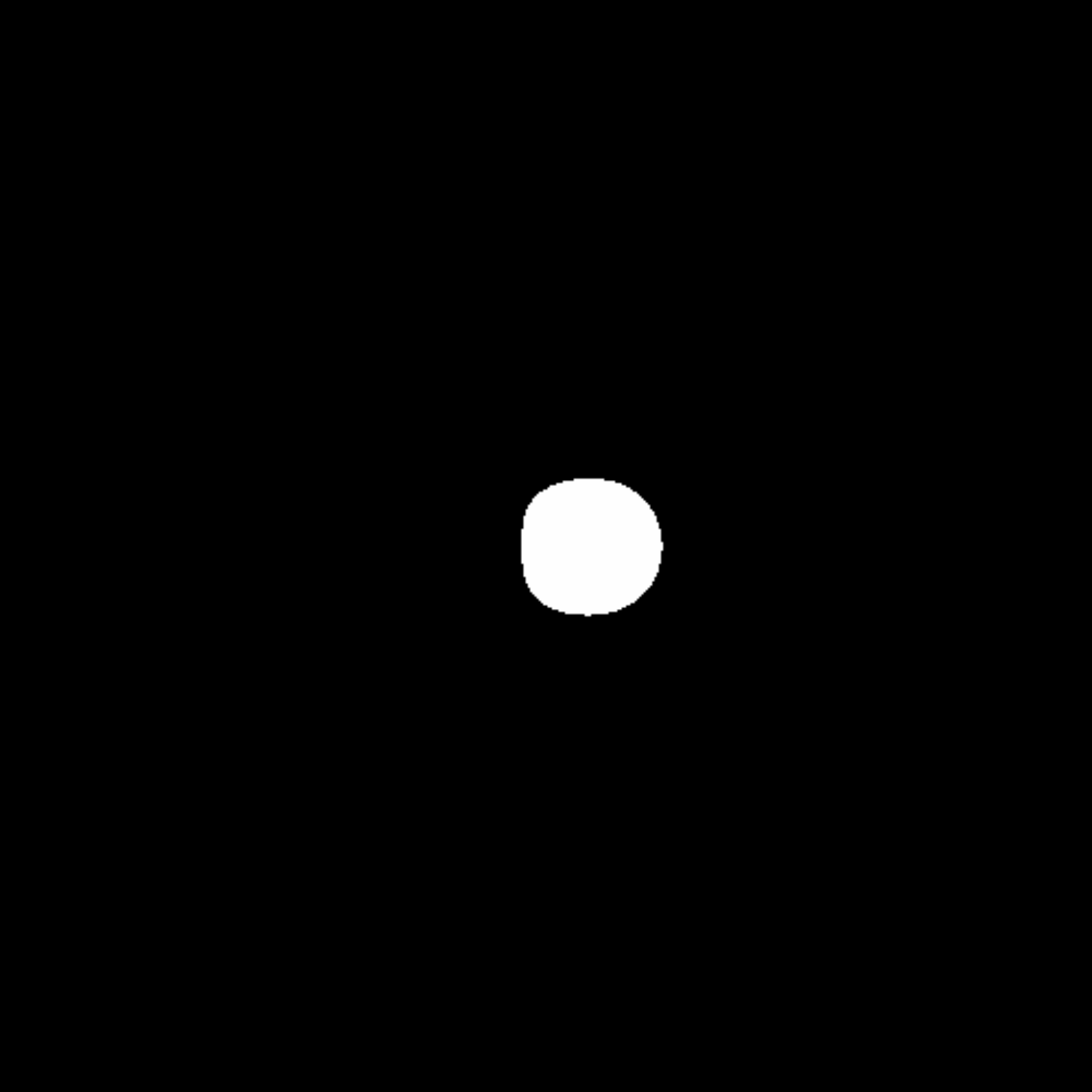}
\includegraphics[width=6cm,height=6cm]{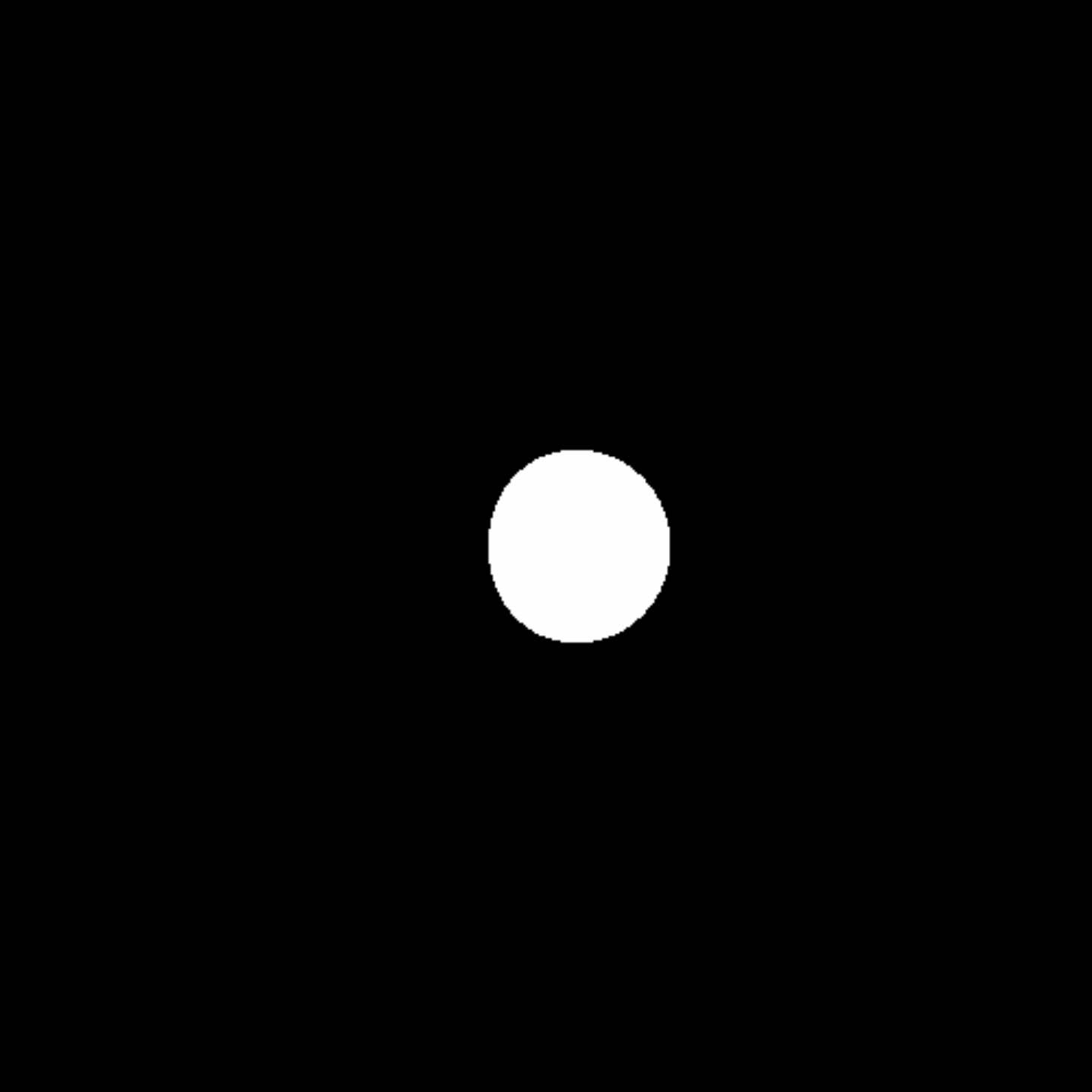}
\caption{\textit{\textbf{Left:} shadow of the \kk of configuration II (field of view $300\,\mu$as).
\textbf{Right:} the same for the comparable Kerr BH ($j_{\mathrm{ADM}}=0.85$).}}
\label{fig:shadow2}
\end{figure}

Fig.~\ref{fig:shadow2} shows the shadows of the \kk and comparable Kerr BH of our configuration II.
They are very similar to the corresponding one in Fig.~5 of~\cite{Cunha:2015yba} (second line from the top).
It is clear that the angular size of the shadow is smaller (by $\approx 25\%$) for the \kk as compared to the comparable Kerr setup.

\begin{figure}[htbp]
\centering
\includegraphics[width=6cm,height=6cm]{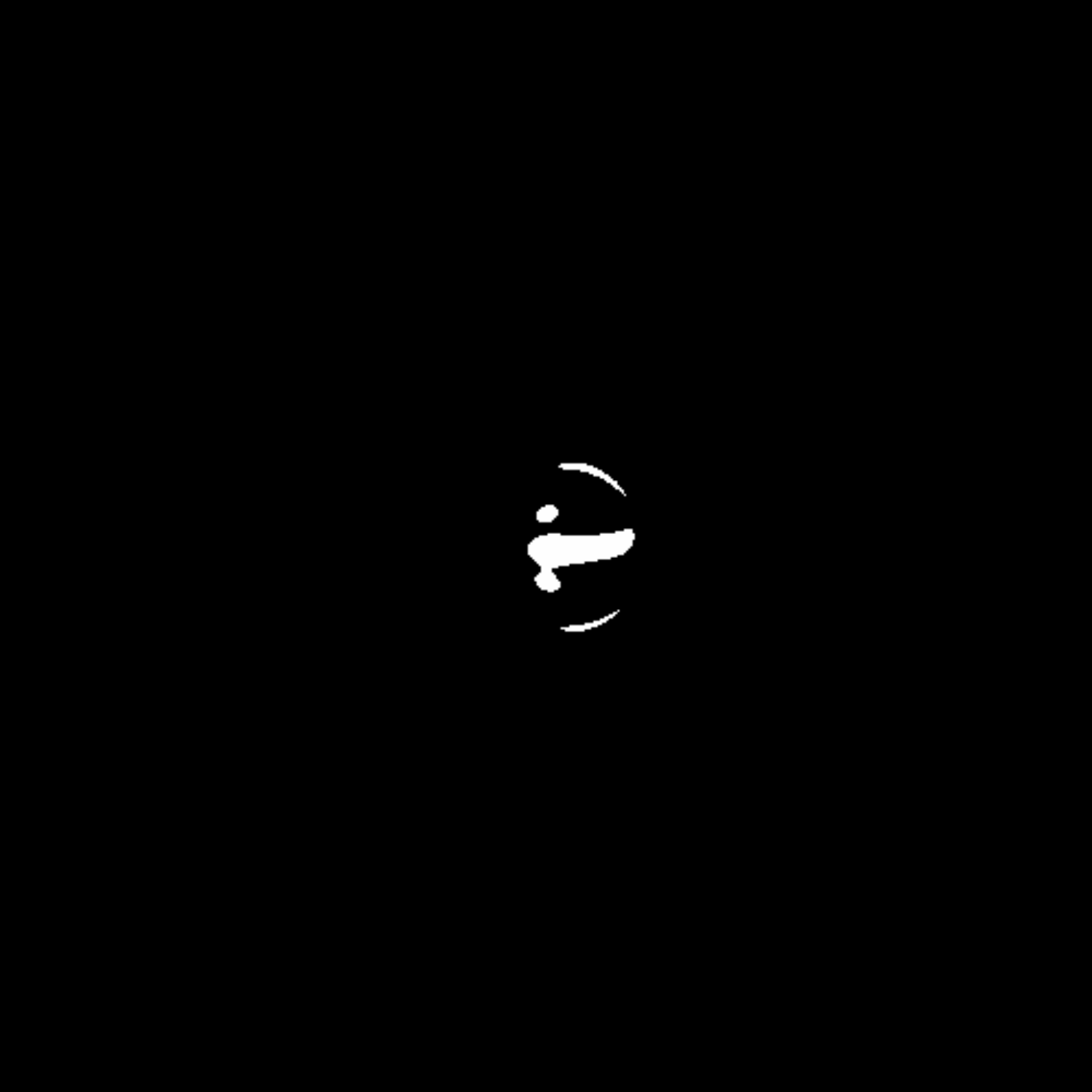}
\includegraphics[width=6cm,height=6cm]{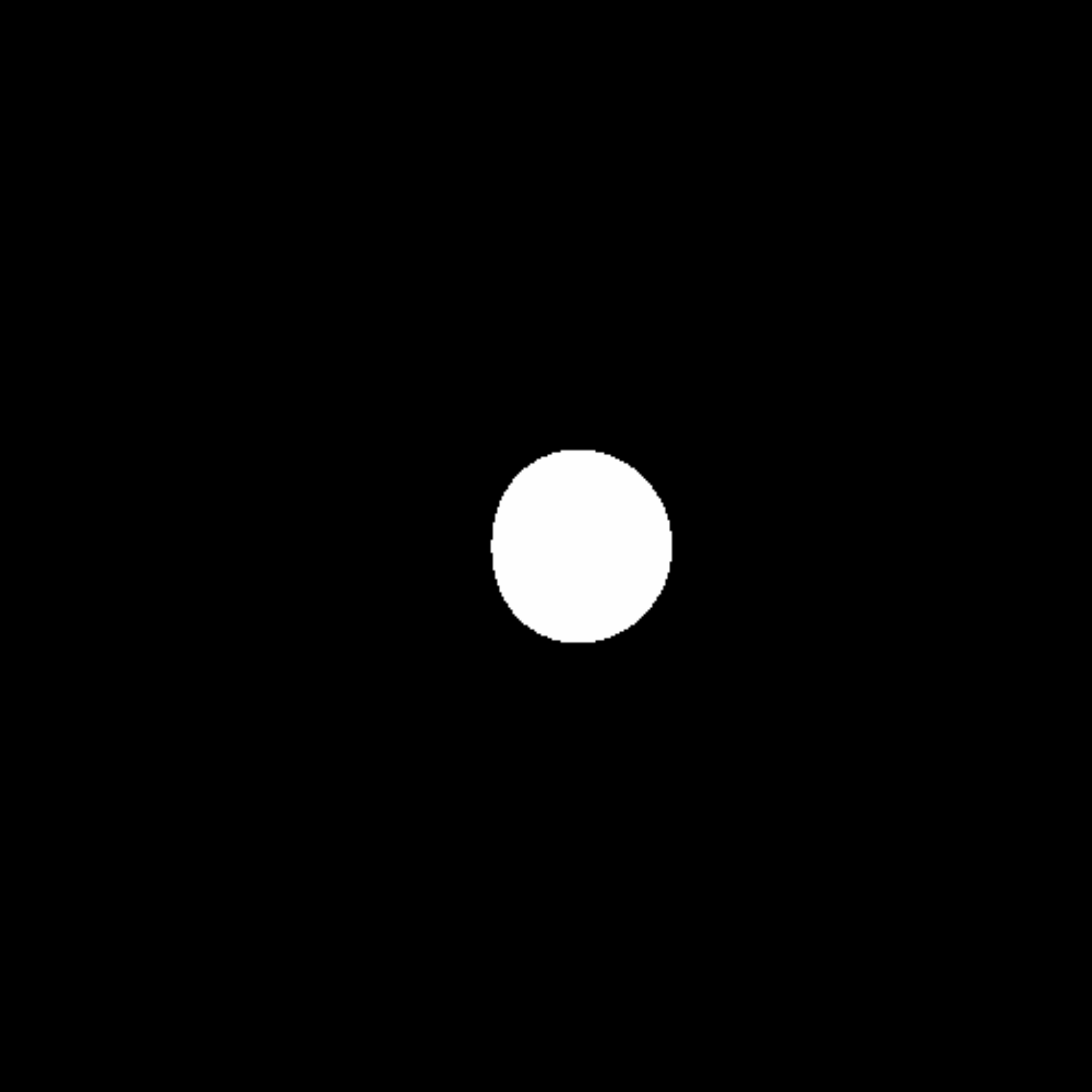}
\caption{\textit{\textbf{Left:} shadow of the \kk of configuration III (field of view $300\,\mu$as).
\textbf{Right:} the same for the comparable Kerr BH ($j_{\mathrm{ADM}}=0.894$).}}
\label{fig:shadow3}
\end{figure}

\begin{figure}[htbp]
\centering
\includegraphics[width=6cm,height=6cm]{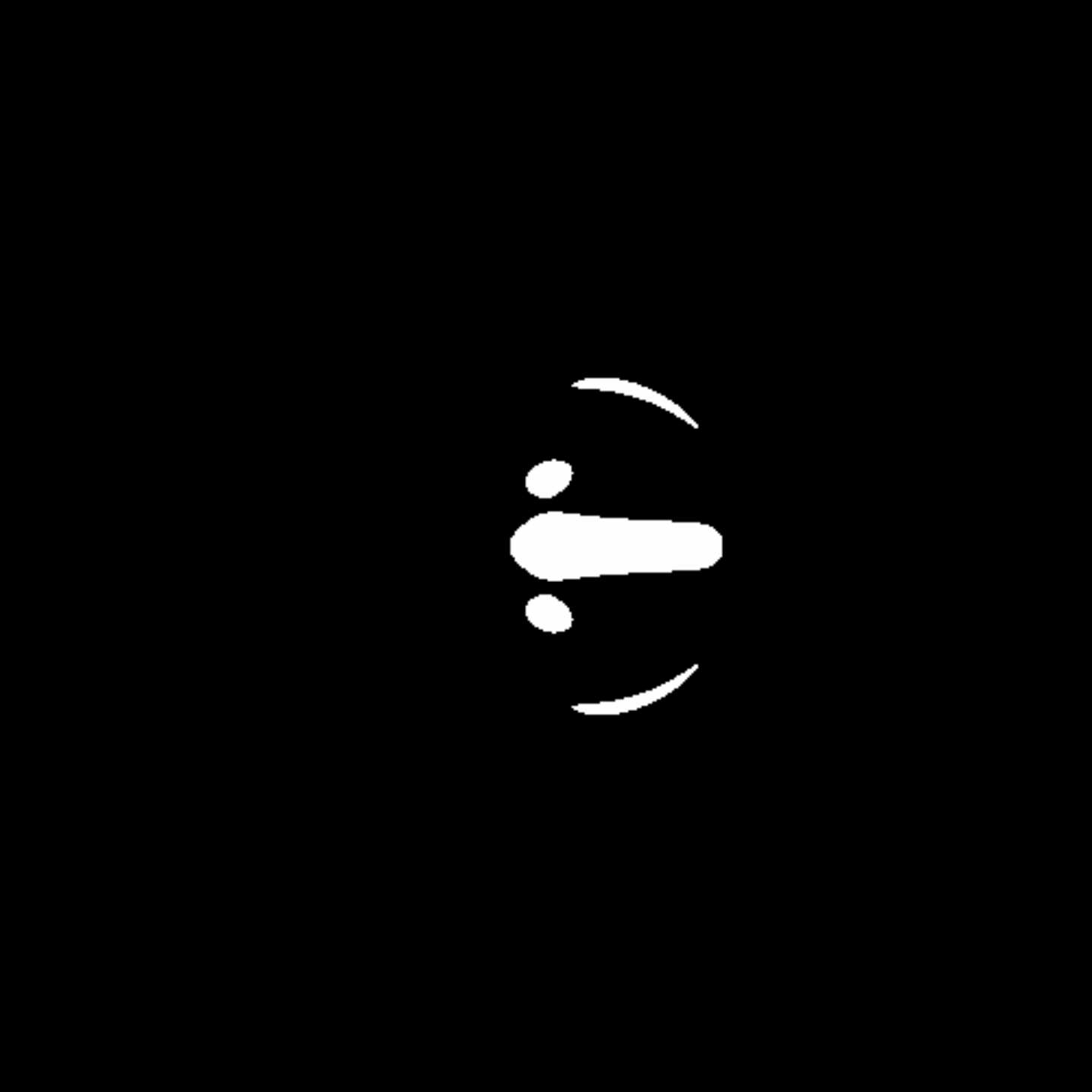}
\includegraphics[width=6cm,height=6cm]{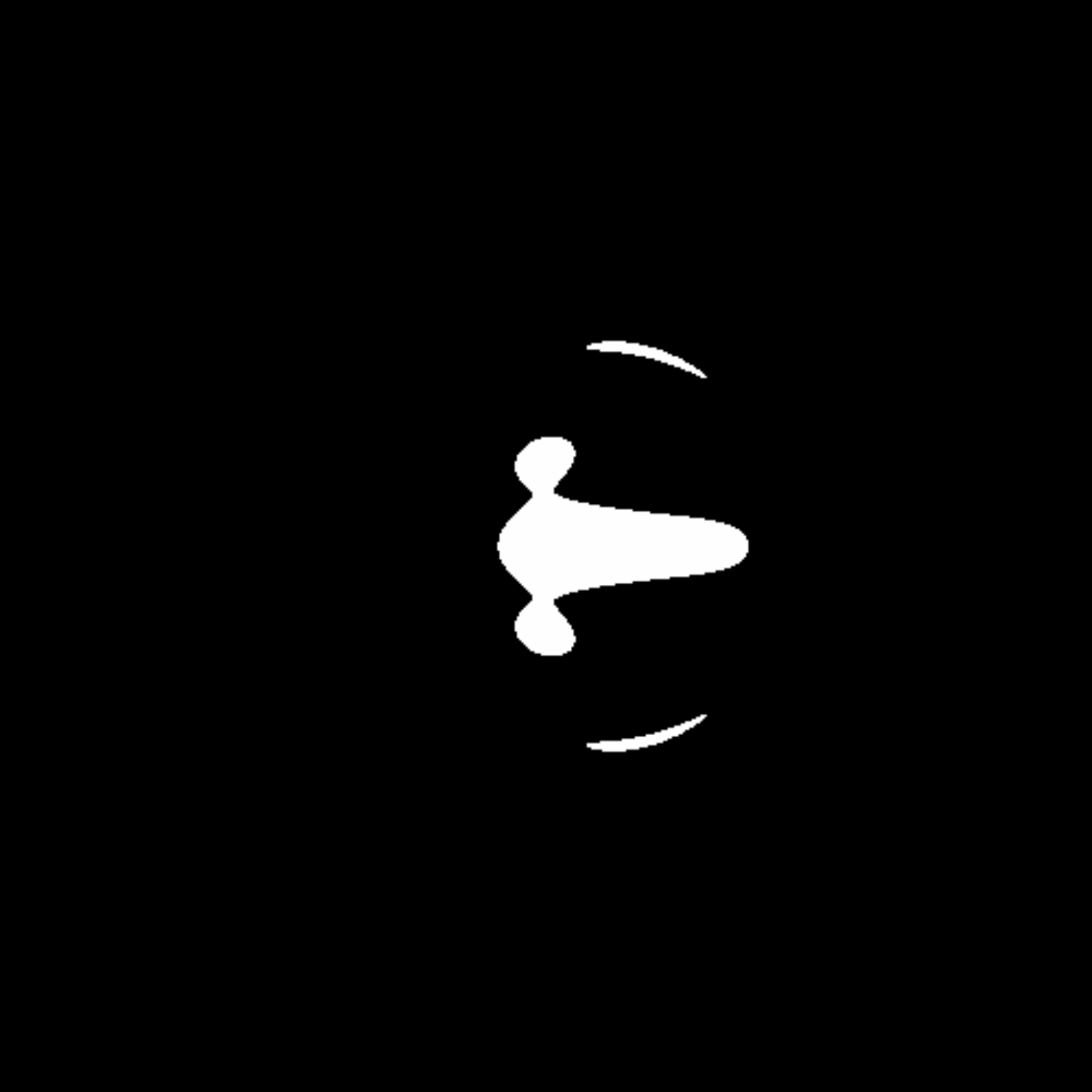}
\caption{\textit{\textbf{Left:} shadow of the \kk of configuration III from an inclination of $\theta=90^\circ$ (field of view $150\,\mu$as).
\textbf{Right:} same computation, changing only the coordinate radius of the observer which is taken
the same as in~\cite{Cunha:2015yba} 
(field of view
$\pi/2$). {Observe that the shadow's topology depends on the observer's distance.}}}
\label{fig:shadow3b}
\end{figure}

Fig.~\ref{fig:shadow3} shows the shadows of the \kk and comparable Kerr BH of configuration III.
The two shadows are extremely different, as already showed in~\cite{Cunha:2015yba}.
The \kk shadow is rather different from the lower-left panel of Fig.~5 of~\cite{Cunha:2015yba}, but
the dominating difference is due to the change in inclination $\theta$ since our
shadow is not symmetric with respect to the horizontal axis.
In order to be able to compare more precisely with~\cite{Cunha:2015yba}, Fig.~\ref{fig:shadow3b} shows the shadow of the \kk of configuration III
as observed from the equatorial plane ($\theta=90^\circ$) and for a smaller field of view ($150\,\mu$as). The left panel shows the shadow as observed from the Earth, while the
right panel shows the shadow as observed by a very close observer, at the same radial coordinate as used in~\cite{Cunha:2015yba}.
These two computations show that the topology of the shadow changes with the radial coordinate of the observer.
The right panel of Fig.~\ref{fig:shadow3b} is extremely similar to the lower-left panel of Fig.~5 of~\cite{Cunha:2015yba}
thus validating the ray-tracing comparison in the most non-Kerr-like spacetime.

\section{Accretion torus and images}
\label{sec4}

We consider a simple toroidal accretion structure surrounding the various BHs
of configurations I, II and III. This model is the same as presented in~\cite{vincent15kerr}
and recently applied to boson stars~\cite{Vincent:2015xta}. It is based on the magnetized
torus model of~\cite{komissarov06}. We will only present this accretion
model very briefly and refer to~\cite{vincent15kerr} for details. The accretion torus model
that we consider is made of a non-self-gravitating perfect polytropic fluid, circularly orbiting with constant specific
angular momentum $\ell = -u_\varphi/u_t$, where $\mathbf{u}$ is the fluid 4-velocity, which
is completely fixed by imposing the constancy of $\ell$ and the circular motion.
The original work of~\cite{komissarov06} considers a toroidal magnetic field. However, \cite{Vincent:2015xta}
showed that the synchrotron images are not sensitive to the direction of the magnetic field.
We thus simplify the original model of~\cite{komissarov06} by considering an isotropized magnetic field.
The energy-momentum conservation is readily integrated by considering a polytropic-like equation of state 
with the gas pressure $p$ and enthalpy $h$ (equal to the sum of the gas pressure and total energy density)
related through
\be
p = \kappa h^k \ ,
\ee
where $k$ is the polytropic exponent, and $\kappa$ is the polytropic constant.
Conservation of energy-momentum then leads to
\be
W_s-W+\frac{k}{k-1}\kappa\, h^{k-1}=0 \ ,
\ee
where $W=-\mathrm{ln}|u_t|$ is a potential (known throughout spacetime
because the 4-velocity is fixed once $\ell$ is chosen) and $W_s$ is its value at the surface of the torus.
This immediately gives
\bea
h &=& h_c \,\omega^{1/(k-1)}\ , \\ \nn
\kappa &=& \left(W_c-W_s\right) \frac{k-1}{k} h_c^{1-k} \ ,
\eea
where $h_c$ is the central enthalpy, $\omega = (W-W_s)/(W_c-W_s)$ and $W_c$ is the potential value
at the center of the torus. Thus, the enthalpy is analytically
known throughout the torus. The values of gas pressure (from the polytropic relation), magnetic pressure ($p_m = p/\beta$, where $\beta$
is a chosen parameter), magnetic field ($B^2 = 24 \pi \,p_m$) and temperature (from the perfect-gas relation) immediately follow.
This torus emits thermal synchrotron radiation, following the prescription given in~\cite{vincent15kerr}.
We note that the term accretion may be misleading given that our model is stationary.
However, we consider this torus as a simple model for an instantaneous snapshot
of a more realistic time-evolving accretion flow, so we keep referring to it as an accretion torus.
The torus model is fully described by the choice of a particular background spacetime
plus the choice of a set of $7$ astrophysical parameters. These are the torus constant angular
momentum $\ell$ and inner radius $r_{\mathrm{in}}$ (fixing these two parameters sets
the outer radius of the torus), the inclination {of the observer} $\theta$, 
the torus
central electron number density $n_c$ and temperature $T_c$, the polytropic exponent $k$ relating
pressure and enthalpy, and the gas-to-magnetic pressure ratio $\beta$. Among these
astrophysical parameters, only $r_\mathrm{in}$ will be varied.
The other are fixed to the values given in Table~\ref{tab:param} and were chosen to
give reasonable values of fluxes as compared to millimeter observed data of Sgr~A*,
and to get a rather compact structure ($i.e.$ not a very extended torus but rather a
structure extending over a small radial distance).
\begin{table}[htbp!]
\centering \caption{Torus model astrophysical parameters (the spacetime is not considered here).
{The value of $r_{\mathrm{in}}$ is slightly varied from spacetime to spacetime to keep the same
angular size of the structure, so only an approximate value is given here.}}
\vspace{0.2cm}
\begin{tabular}{l{c}c}
\hline
parameter                  &             & value                                \\
\hline
inner radius  			&  	$r_{\mathrm{in}}$		 &  $\approx 5.5\,M$   \\
angular momentum           & $\ell$   &               $3.6 \, M$                    \\
inclination                & $\theta$         &      $85^\circ$          \\
central density ($ \mathrm{cm}^{-3}$)           & $n_c$       &        $6.3\times10^6$                        \\
central electron temperature (K)            & $T_c$   &     $5.3\times10^{10}$  \\
polytropic exponent           & $k$           & $5/3$                                     \\
gas to magnetic pressure ratio              & $\beta$ & $10$                 \\
\end{tabular}
\label{tab:param}
\end{table}
For one given spacetime, the inner radius is fixed such that the (SP) radial
extent of the structure (outer radius minus inner radius) is close to $20\,M$.
This ensures to obtain an emitting structure with an angular size satisfying the constraint
imposed by the first EHT data~\cite{doeleman08}. Thus, we choose to vary $r_\mathrm{in}$
from spacetime to spacetime in order to maintain an approximately constant angular size
of the structure as observed from Earth. Note, however, that the range of variation of $r_\mathrm{in}$
is very small and that this quantity always stays close to $r_\mathrm{in} \approx 5.5\,M$.

The \textsc{Gyoto} code allows to ray trace photons from a distant observer and integrate
the radiative transfer equation through the optically thin synchrotron-emitting accretion torus,
thus producing a map of specific intensity, $i.e.$ an image. We consider an observed frequency
of $230$~GHz for the ray-traced photons, corresponding to the frequency used for the
early EHT science~\cite{doeleman08}.
The aim of this section is to compare each
\kks image to the image of the comparable Kerr configuration, i.e. that having
the same ADM mass $M$ and the same total angular momentum $J$ (cf. Sec.~\ref{sec2}).

\begin{figure}[htbp]
\centering
\includegraphics[width=7cm,height=7.4cm]{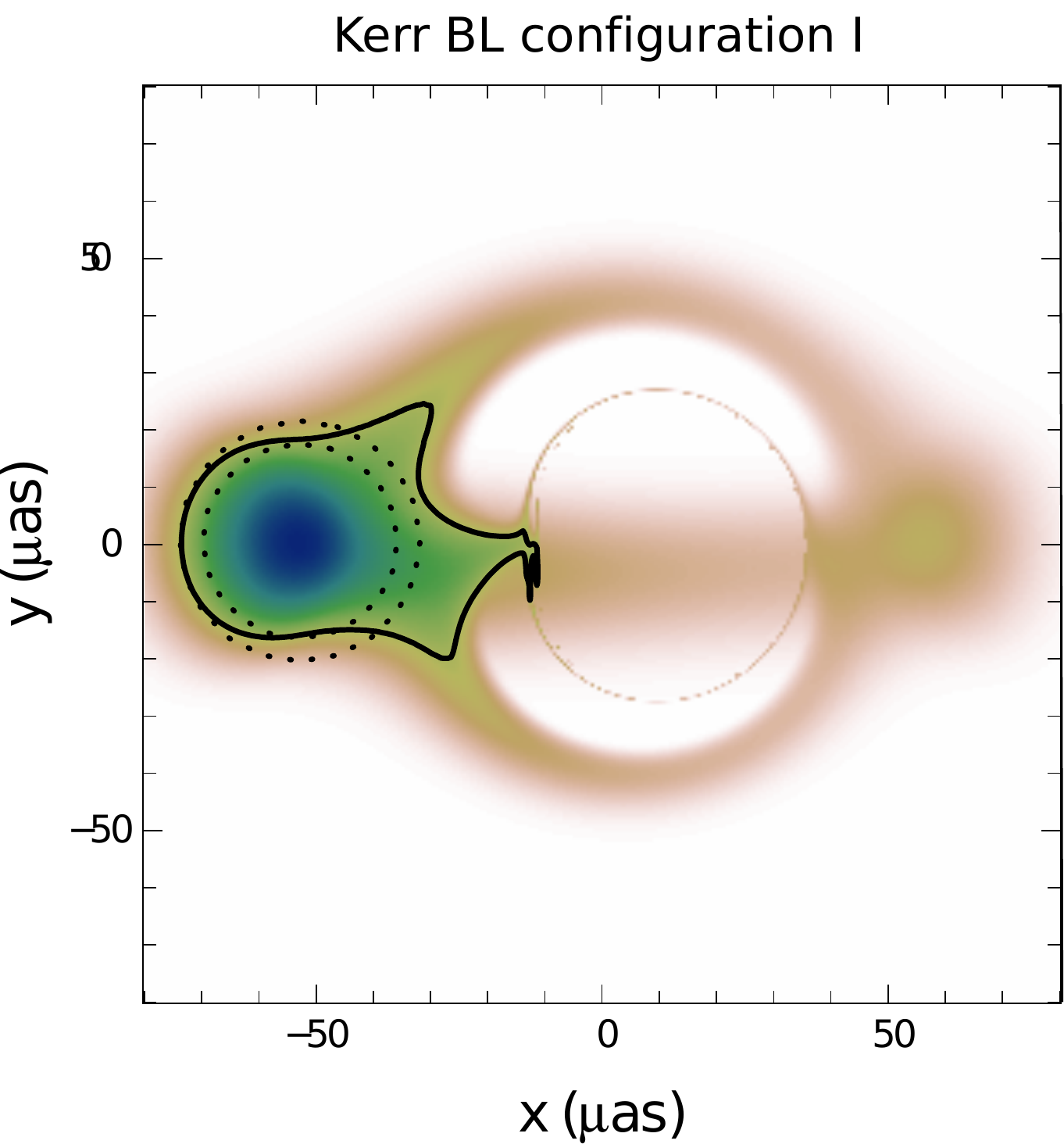} \\
\includegraphics[width=7cm,height=7.4cm]{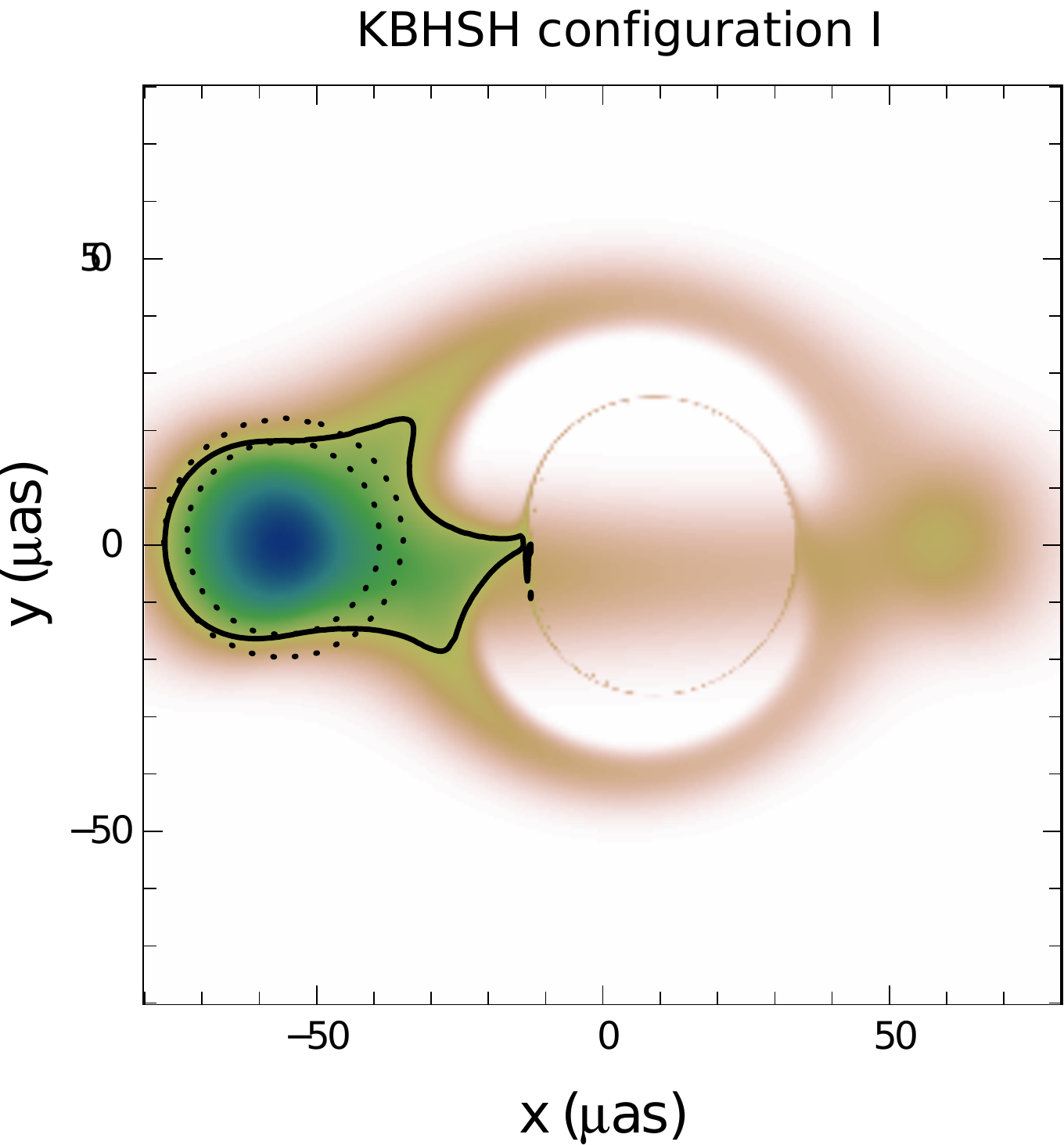}
\includegraphics[width=7cm,height=7.4cm]{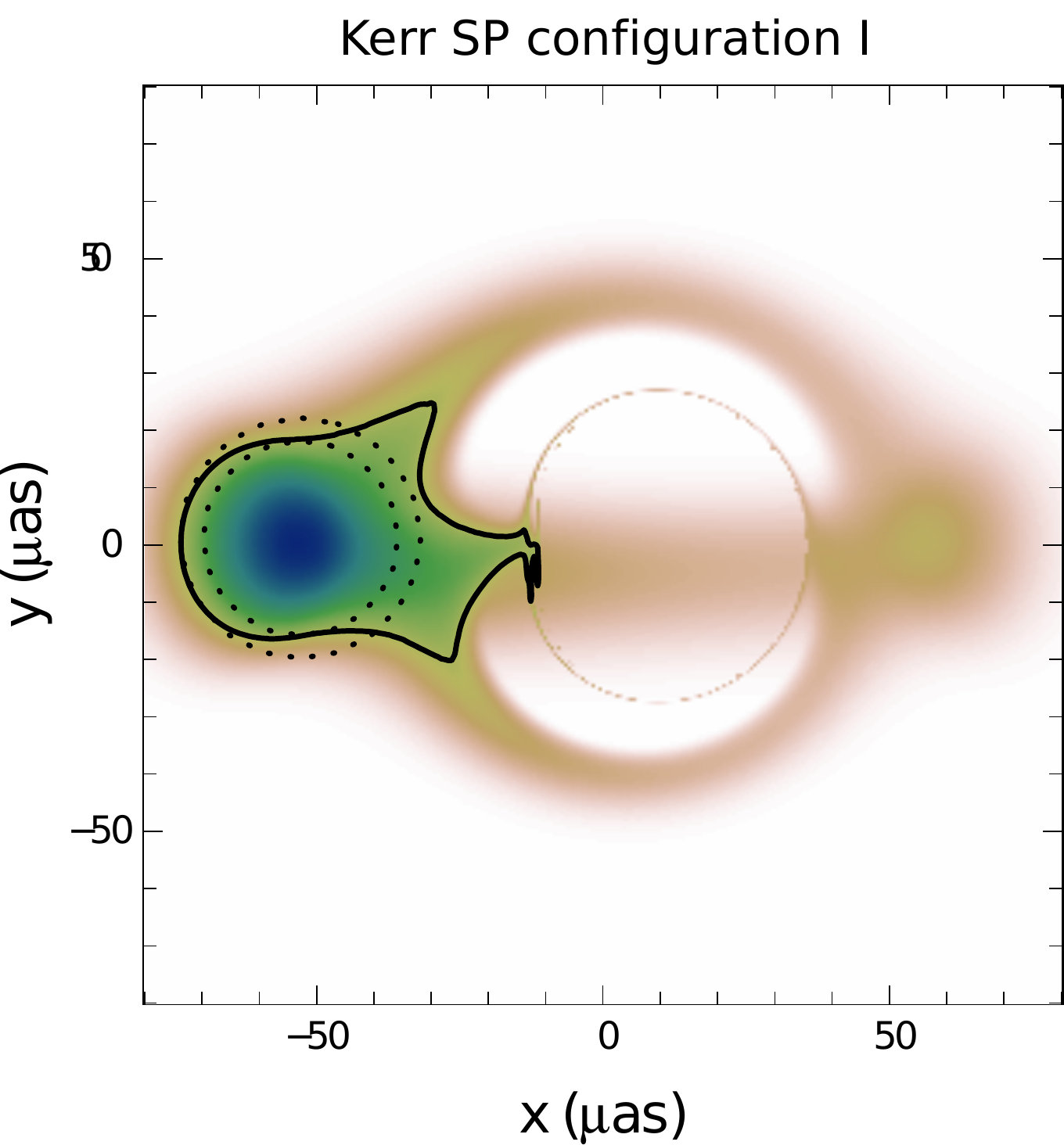}
\caption{\textit{\textbf{Configuration I images.} \textbf{Up:} image at $230$~GHz of the torus model surrounding the Kerr BH of configuration I,
computed with an analytical Boyer-Lindquist (BL) metric. {All images in this article are represented in inverse colors: high intensity is in dark blue,
low intensity is in yellow/orange.}
The dotted circles show the $1\sigma$ upper and lower confidence limits for the intrinsic angular size of the emitting zone~\cite{doeleman08}. The solid black contour encompasses the region of the accretion flow emitting 50$\%$ of the total flux: it is considered as an order-zero approximation of the size of the emitting region. The image has thus a reasonable angular size provided that the solid contour approximately lies within the dotted circles, which is the case here and in all other figures of the article.
\textbf{Lower left:} the corresponding \kk setup (same ADM mass and same angular
momentum).
\textbf{Lower right:} the same image as in the upper panel, but using a numerical Kerr spacetime described with SP coordinates.
}}
\label{fig:imageConf1}
\end{figure}

As a first check, we have verified that
Kerr BH images computed with SP coordinates were indistinguishable from
Kerr images computed with the more standard Boyer-Lindquist (BL) coordinates. Fig.~\ref{fig:imageConf1}
(upper and lower-right panels) shows that this is indeed so:
the comparable Kerr image of configuration I computed {using the numerical solution} in SP coordinates differs
by only $\approx 0.7 \,\%$ with respect to the same image computed {using the analytical solution} in Boyer-Lindquist coordinates.

The lower-left  and upper panels of Fig.~\ref{fig:imageConf1} show the ray-traced images of the \kk and comparable Kerr BH
of configuration I.
The \kk and Kerr images are very similar, which is not surprising given that the \kk of configuration I
was chosen to be very Kerr-like.
The flux difference between the \kk and Kerr images is $\approx 0.04\%$, which is vanishingly small
as far as spectral observations are concerned.
It may seem surprising that this flux ratio is actually smaller than the ratio between the
analytical Kerr BL and numerical Kerr SP spacetimes discussed above. This fact is explained by two causes.
First, the Kerr BL and Kerr SP spacetimes use different
coordinates. This introduces a numerical error that is not present when comparing Kerr SP and \kk spacetimes
that use the same coordinates. Second, the average specific intensity ratio, when comparing pixel by pixel,
is of $0.5\%$ for the Kerr BL/SP comparison and $4.5\%$ for the Kerr SP/\kk comparison. The flux ratio
is different, because most of the flux actually comes from the center of the torus, and in these regions, the
numerical error due to the change of coordinates between Kerr BL and SP dominates over the difference
between the Kerr SP and \kk spacetimes.
Arguably the most interesting feature in such strong-field
image is the thin ring of illuminated pixels at the center of the image, the photon ring. The \kk and Kerr
photon rings are very similar in shape, but their sizes are differing by $\approx 5 \%$. The recent study
of~\cite{psaltis15} discusses the current measured error on the ratio $M/D$ of the mass of Sgr~A* over its distance,
and shows that it is of $\approx 6 \%$. Moreover, this study advocates that EHT data could lead to
a constraint of the photon ring angular size (the most advanced goal of the EHT) with a precision of $\approx 10\%$. The size difference of $5\%$ of the
\kk and Kerr photon rings that we report here is thus most probably unobservable.

\begin{figure}[htbp]
\centering
\includegraphics[width=7cm,height=7.4cm]{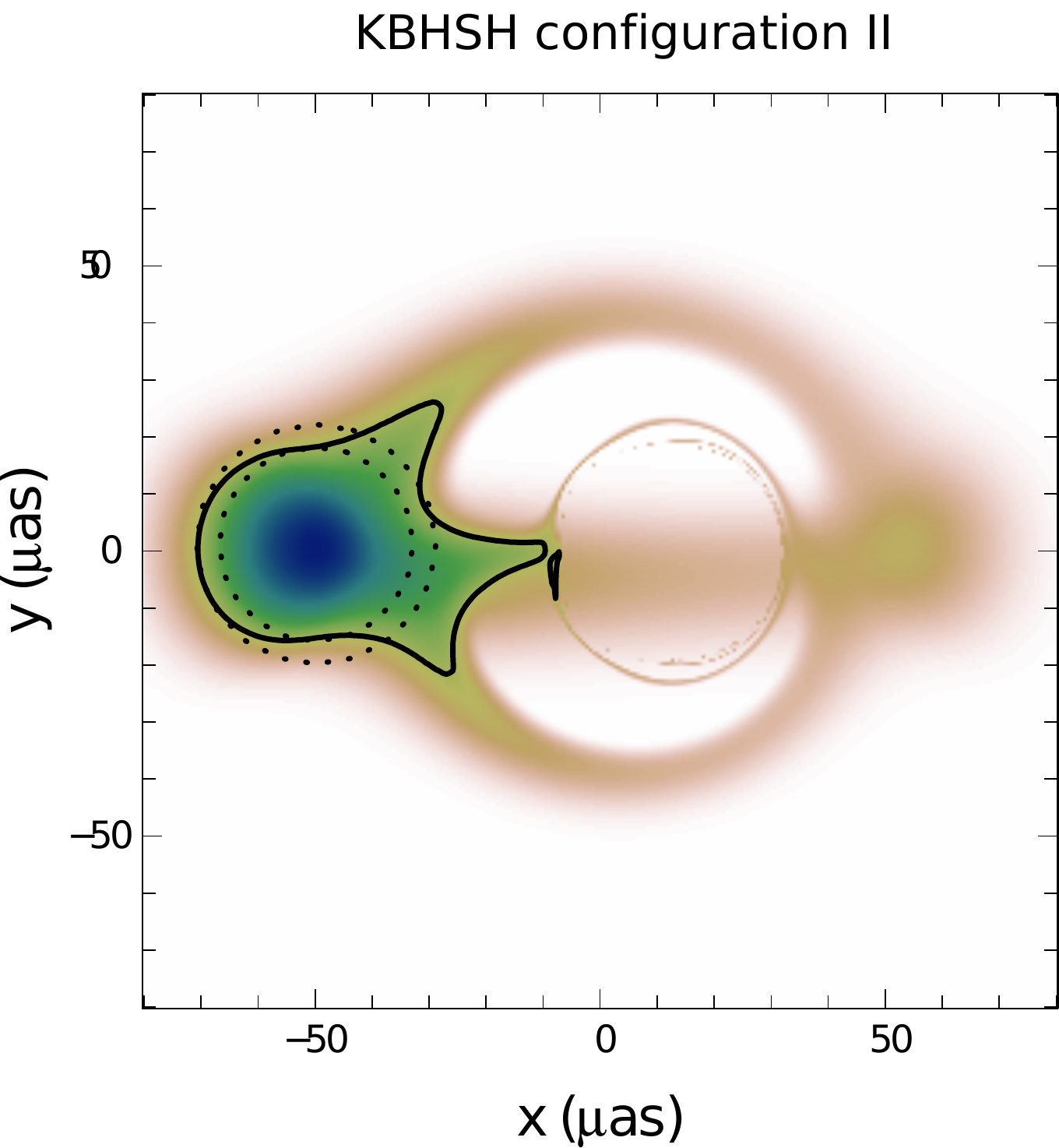}
\includegraphics[width=7cm,height=7.4cm]{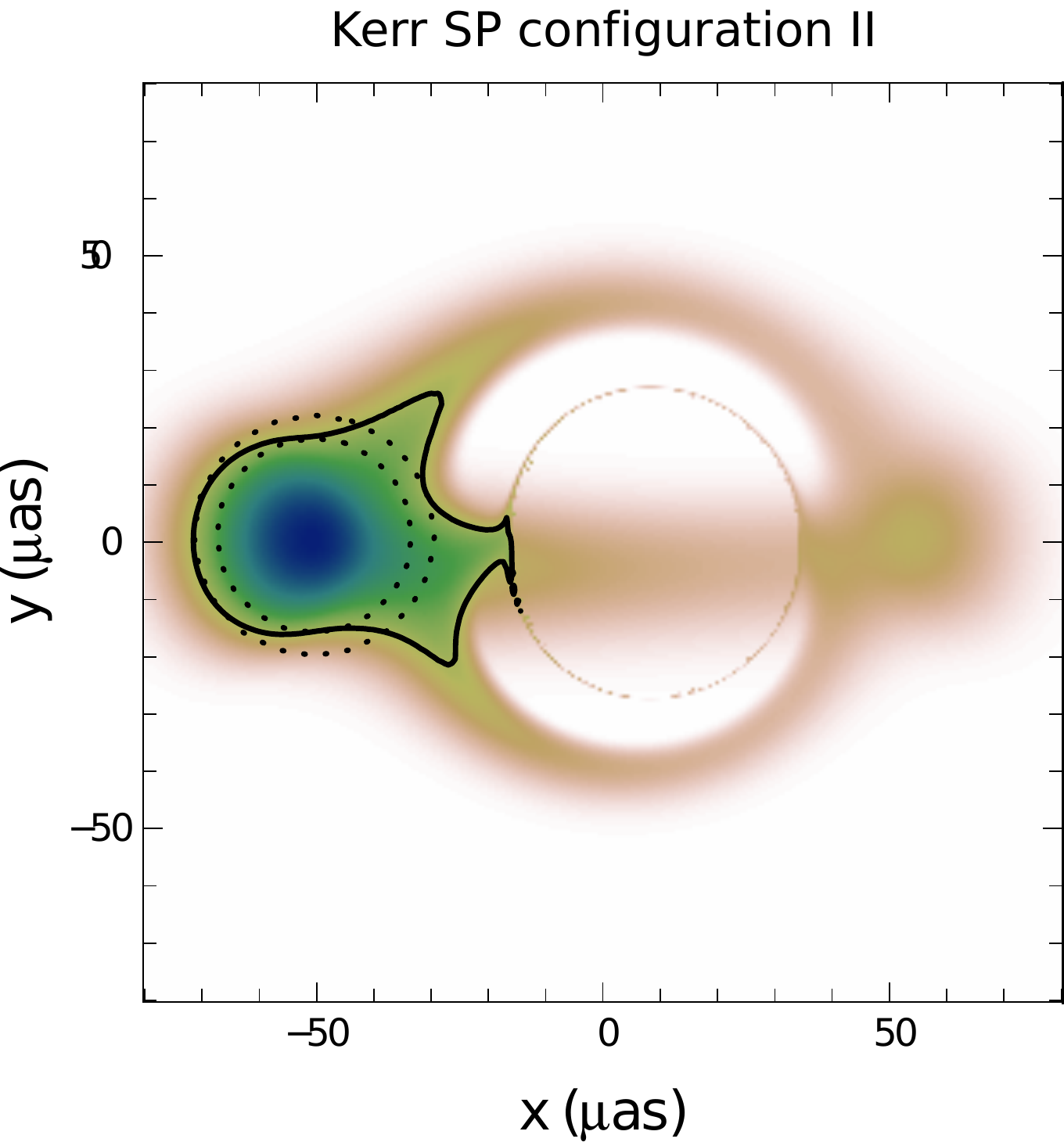}
\caption{\textit{\textbf{Configuration II images.} Same as in Fig.~\ref{fig:imageConf1}. \textbf{Left:} Image of an accretion torus
surrounding the \kk of configuration II.
\textbf{Right:} Same image for the comparable Kerr case.}}
\label{fig:imageConf2}
\end{figure}

\begin{figure}[htbp]
\centering
\includegraphics[width=7cm,height=7.4cm]{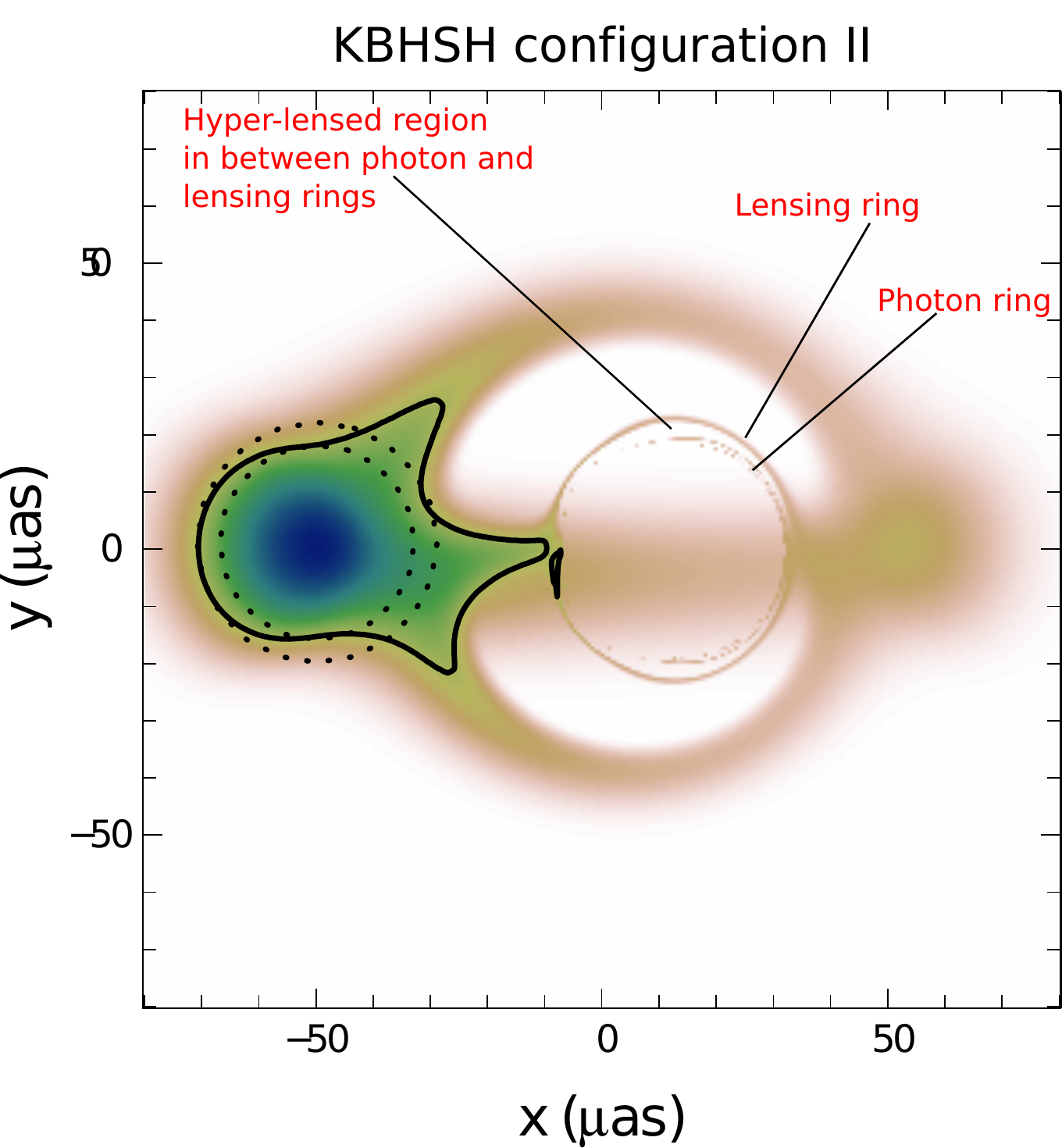}
\caption{\textit{Illustration of the various interesting regions in the image: the lensing
ring is the outermost ring, it is the outer boundary of the hyper-lensed region. {The left
part of the lensing ring is visibly distorted with respect to a Kerr photon ring}. The inner
ring is the photon ring. It is both the inner boundary of the hyper-lensed region and the
outer boundary of the shadow.}}
\label{fig:defregion}
\end{figure}

Fig.~\ref{fig:imageConf2} shows the images of the accretion torus surrounding the \kk and comparable
Kerr solution of configuration II.
The flux difference between the \kk and Kerr images is of $\approx 1.5\,\%$, which is still very small
with respect to the error bars of spectral observations at $230$~GHz.
The first difference of the \kk image with respect to Kerr is the fact that there are two edges in the intensity
distribution: there is a brighter ring at the center of the image that looks like a distorted Kerr photon ring  {(the left part is visibly different from a portion of a circle)},
and inside this first ring, a second, fainter one. The interpretation of these two features is helped by
considering the shadows in Fig.~5 of~\cite{Cunha:2015yba}. These images show clearly that as
the spacetime becomes more non-Kerr-like, a region develops around the shadow that is affected
by intense and complex lensing effects (this is particularly clear in the third row from the top in Fig.~5
of~\cite{Cunha:2015yba}). In the following we will call this region the \textit{hyper-lensed}
region, and its outer boundary the \textit{lensing ring} (see Fig.~\ref{fig:defregion}).
The brighter ring in the left panel of our Fig.~\ref{fig:imageConf2} is this lensing
ring, while the fainter ring is the photon ring. We have checked that photons
forming the photon ring approach closer to the event horizon (in radial coordinate)
than photons forming the lensing ring. This is the expected behavior given that the
photon ring is the projection of the innermost photon orbit and marks the innermost
limit a photon can visit without falling into the event horizon.
It is very probable that EHT would detect
only the brighter lensing ring, so we will focus on it in the following.
The \kk lensing ring is distorted, in the sense that it does not look like the photon ring of any Kerr
BH (see~\cite{chan13} for an overview of Kerr photon rings). However, this distortion
is extremely tiny and would most probably be unnoticed by EHT observations. Still, the \kk lensing ring
is smaller in angular size, as compared to the Kerr photon ring.
The difference in angular size between the \kk lensing ring and Kerr photon ring reaches $20\,\%$, which
is bigger than the error on the ratio $M/D$ as discussed above, and also bigger than the foreseen
precision of the EHT data for constraining the photon ring angular size of Sgr~A*.
Consequently, should the EHT be capable of giving a constraint
on the angular size of the photon ring of Sgr~A* to within $\approx 10\%$ and should this value be
too small to be compatible with a Kerr BH of mass $M$ at distance $D$, this would support
the existence of an alternative compact object such as a \kk at the Galactic center. Note that the
comparable Kerr BH of configuration II has a spin of $j_{\mathrm{ADM}} = 0.85$. A Kerr
BH with spin $j_{\mathrm{ADM}} = 0.999$ would have a photon ring smaller by approximately $3\%$~\cite{chan13},
so the difference of $20\%$ between the Kerr and \kk ring angular sizes cannot be accounted for
by varying the Kerr BH spin.

\begin{figure}[htbp]
\centering
\includegraphics[width=7cm,height=7.4cm]{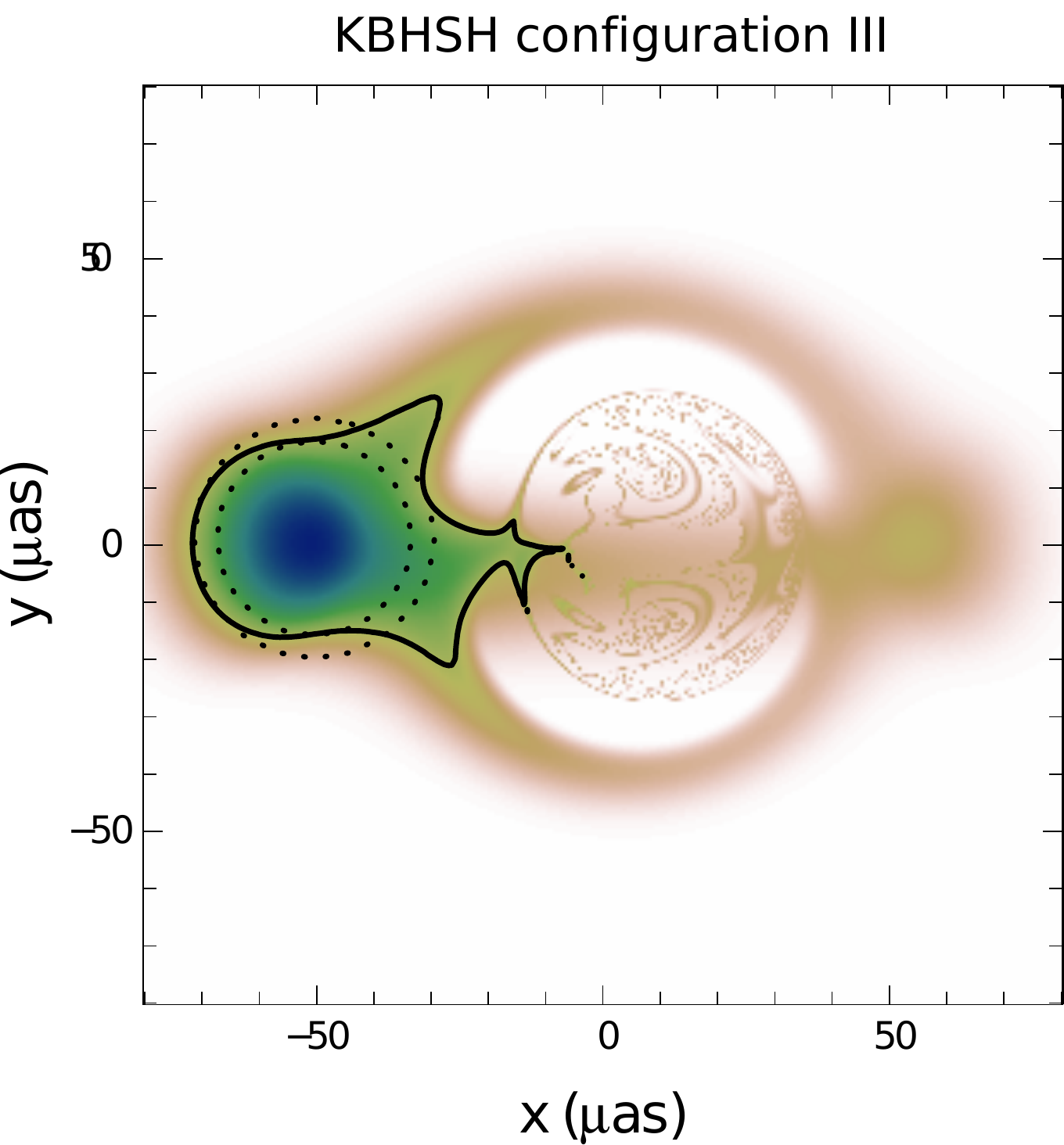}
\includegraphics[width=7cm,height=7.4cm]{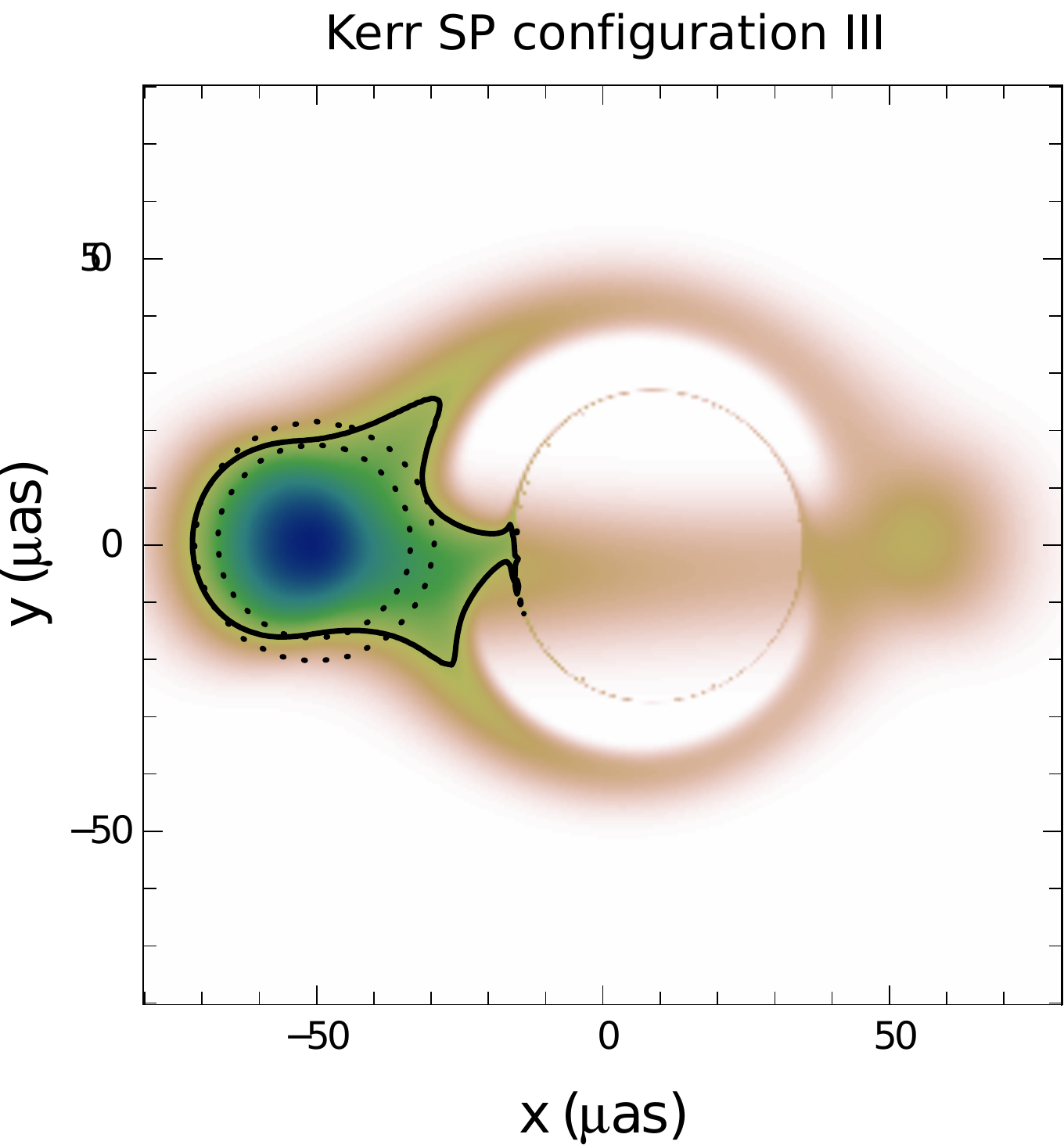}
\caption{\textit{\textbf{Configuration III images.} Same as in Fig.~\ref{fig:imageConf1}. \textbf{Left:} Image of an accretion torus
surrounding the \kk of configuration III. Note that the outer boundary of the central "noisy" region is not a photon ring,
it is a lensing ring (see text for details).
\textbf{Right:} Same image for the comparable Kerr case. The Kerr photon ring is very similar to the
\kk lensing ring.}}
\label{fig:imageConf3}
\end{figure}

Finally, Fig.~\ref{fig:imageConf3} shows images of our accretion torus model in
the \kk and comparable Kerr spacetimes of configuration III.
The flux difference between the two images is of $\approx 4\%$, still smaller than the observational error bars.
The most striking feature of the left panel in this image is the central ``noisy" region, full of radiation.
This is the hyper-lensed region. Its outer boundary, the lensing ring,
nearly coincides with the comparable Kerr photon ring, and is bigger and less distorted than the lensing ring
of the \kk of configuration II. This is in agreement with the findings of~\cite{Cunha:2015yba}.
Why is the hyper-lensed region of the configuration III \kk ``noisy" while the hyper-lensed region
of its configuration II counterpart is empty of radiation? Because of the very small angular size of the
hyper-lensed region in configuration II. Photons from inside the hyper-lensed region will carry
a non-negligible amount of radiation provided they visit the central regions of the accretion torus
where most of the radiation is produced. In configuration II, no photon forming the inside of the hyper-lensed region
visit these central parts of the torus. Only photons forming the lensing ring do so, because the lensing ring
corresponds to extremely bent photons and a large region of spacetime (including the inner parts of the torus)
is projected to this thin ring of pixels, which is thus bright.
On the contrary, many photons from the much bigger hyper-lensed region of the \kk of
configuration III visit the central parts of the torus thus leading to a lot of radiation
being located in this region of the image.
The BH shadow is not visible in the left panel of our Fig.~\ref{fig:imageConf3}.
It is actually so small in angular size (see Fig.~\ref{fig:shadow3}) that it is nearly completely erased by the radiation emitted by the part of the
torus located in between the BH and the observer. The photon ring of this spacetime, being the
outer boundary of the shadow, is invisible.
In this spacetime, the important features, as far as EHT observations
are concerned, are the hyper-lensed region and lensing ring. The shadow and photon ring are not interesting
observation-wise.

The hyper-lensed region with a large angular scale is a very interesting feature of the \kk spacetime of configuration III because
it may also lead to observational difference, for different reasons than for the \kk of configuration II.
The obvious difference between the two panels of Fig.~\ref{fig:imageConf3} is the fact that there
is flux all around the hyper-lensed region in the \kk spacetime, while the shadow of the Kerr spacetime
is free of radiation, except for the radiation emitted in the foreground by the part of the torus in between the BH
and the observer. In the \kk spacetime, $\approx 13\,\%$ of the total flux is located in the hyper-lensed region.
In the Kerr spacetime, $\approx 10 \, \%$ of the total flux is located in the shadow, due to emission in the
foreground. Thus, approximately $3\%$ of the total flux of the \kk image of configuration III is a ``hairy flux",
$i.e.$ due to the hyper-lensing effects specific to the \kk spacetime. It is probable that an algorithm (see $e.g.$~\cite{psaltis15,johannsen16}) trying to
detect a Kerr BH shadow on an image similar to the left panel of Fig.~\ref{fig:imageConf3} would not
converge because there is nowhere in the image a photon-ring-like structure:
there is no edge ($i.e.$ strong and localized gradient) in the intensity distribution.

\section{Final Remarks}
\label{sec5}

The perspective of the near-future EHT observations of Sgr~A* makes it very timely to study the observable counterparts
of compact objects alternative to the Kerr BH. Among the many such objects,
\kks are particularly interesting because (1) they are exact solutions of Einstein
field equations, (2) they only necessitate the addition of a scalar field,
a rather ubiquitous object in theoretical physics, with the Higgs boson being
an example of a fundamental scalar field in Nature, (3) they do not imply adding any
astrophysically unclear elements (like the thin shell of gravastars~\cite{mazur04}).

This article shows that \kks might be observationally differentiated with respect to Kerr BHs
by using EHT observations. A too Kerr-like \kk would obviously be impossible to differentiate, as our
configuration I illustrates. However, for sufficiently non-Kerr-like \kk, we have highlighted two
features that may allow making an observational difference. The first such feature, illustrated by
our configuration II, is linked to the angular size of the photon/lensing rings.
The lensing ring is a specific feature to \kks spacetimes, defined in Section~\ref{sec4}, which would observationally be
interpreted as a Kerr photon ring.
A sufficiently non-Kerr-like \kk of the same mass and spin
as a Kerr BH has a lensing ring smaller in angular size than the photon ring of the comparable Kerr image.
This size difference reaches $\approx 20\%$ in
our configuration II, which is sufficient to be detectable by EHT data and to be non degenerate with a Kerr BH.
The increasing non-Kerrness of a \kk spacetime is accompanied
by the development of the central hyper-lensed region (also defined in Section~\ref{sec4}),
the outer boundary of which is the lensing ring.
This hyper-lensed region is the second feature that may allow differentiating \kks from Kerr BHs.
Its outer boundary is rather close to the photon ring of a comparable Kerr BH
for very non-Kerr-like \kks. As a consequence,
the shadow region of a Kerr spacetime, which is characterized by an edge in the intensity distribution,
is replaced by a central ``noisy" region without an edge. This edge in intensity distribution being the
signal that algorithms investigating such images look for, it is likely that such algorithms would fail
finding a shadow region in a \kk spacetime such as our configuration III.

Finally, whereas KBHsSH are certainly an interesting theoretical model for phenomenological deviations from the Kerr paradigm,
one may wonder about their realization as astrophysical objects. An obvious necessary condition is the existence of appropriate scalar fields in Nature, as discussed in the Introduction, either as fundamental fields, or, eventually, as a coarse-graining of more fundamental degrees of freedom.  Another central point is the dynamical stability of these configurations. In this respect, we would like to stress that the vacuum Kerr BH (with no scalar field excited) is an \textit{unstable solution} of the model~\eqref{actionscalar}.~\footnote{The instability of Kerr BHs against low frequency modes of bosonic fields was first discussed by Press and Teukolsky in the setup of a Kerr BH surrounded by a mirror~\cite{Press:1972zz}. Subsequently, Damour et al. found that the confining mechanism provided by the mirror is naturally present if the bosonic field is massive~\cite{Damour:1976kh}. See~\cite{Brito:2015oca} for an overview of superradiant instabilities.} Low frequency scalar modes trapped in the vicinity of the BH trigger a superradiant instability, that grows hair around the BH. Recently, the non-linear development of this superradiant instability was shown, in toy models where electric charge is taken as a surrogate for rotation, to lead to a hairy BH~\cite{Sanchis-Gual:2015lje,Bosch:2016vcp}. What really happens in the Kerr case is still an open issue, but it seems plausible that some KBHsSH play a role, either as long-lived transient states, or even as final states in the non-linear development of this instability. A reasonable expectation is that there are different types of (in)stabilities in the full domain of existence of KBHsSH, with different decay timescales. Efforts to understand these issues in detail are currently underway.

\section*{Acknowledgements}
We would like to thank P. Cunha and H. R\'unarsson for discussions. C. H. and E. R. acknowledge funding from the FCT-IF programme. This work was partially supported by  the  H2020-MSCA-RISE-2015 Grant No.  StronGrHEP-690904, by the CIDMA project UID/MAT/04106/2013, and by the ANR Grant No. 12-BS01-012-01.

\bibliographystyle{h-physrev4}
\bibliography{HBH}

\end{document}